\documentstyle[aps,preprint]{revtex}

\begin{document}

\title{Allowed Gamow-Teller Excitations from the Ground State of $^{14}N$}

\author{S. Aroua$^1$, P. Navr\'{a}til$^2$, L. Zamick$^3$, M.S. Fayache$^1$, 
B. R. Barrett$^4$, J.P. Vary$^5$, N. Smirnova$^6$ and K. Heyde$^6$\\
(1) D\'{e}partement de Physique, Facult\'{e} des Sciences de Tunis,\\
Universit\'{e} de Tunis El-Manar, Tunis 1060, Tunisia\\
\noindent(2) Lawrence Livermore National Laboratory, L-414,\\
P. O. Box 808, Livermore, CA 94551  \\
\noindent(3) Department of Physics and Astronomy, Rutgers University\\
Piscataway, New Jersey 08855\\
\noindent (4) Department of Physics, P.O. Box 210081, University of Arizona, \\
Tuscon, Arizona 85721.\\
\noindent (5) Department of Physics and Astronomy, Iowa State University\\
Ames, Iowa 50011\\
\noindent (6) Vakgroep Subatomaire en Stralingsfysica, University of Gent\\
Proeftuinstraat, 86 B-9000 Gent, Belgium\\
}
\date{\today}
\maketitle

\begin{abstract}
Motivated by the proposed experiment $^{14}N(d,{^2He})^{14}C$, we study the 
final states which can be reached via the allowed Gamow-Teller mechanism. Much 
emphasis has been given in the past to the fact that the transition matrix 
element from the $J^{\pi}=1^+~T=0$ ground state of $^{14}N$ to the 
$J^{\pi}=0^+~T=1$ ground state of $^{14}C$ is very close to zero, despite the 
fact that all the quantum numbers are right for an allowed transition. 
We discuss this problem, but, in particular, focus on the excitations to 
final states with angular momenta $1^+$ and $2^+$. We note that the summed 
strength to the $J^{\pi}=2^+~T=1$ states, calculated with a wide variety of 
interactions, is significantly larger than that to the $J^{\pi}=1^+~T=1$ 
final states.
\end{abstract}

\section{Introduction}

\indent Much attention has been given over the past several decades to the 
fact that the Gamow-Teller ($GT$) matrix element between the $J^{\pi}=1^+~T=0$ 
ground state of $^{14}N$ and the $J^{\pi}=0^+~T=1$ ground state of $^{14}C$ 
(or that of its mirror nucleus $^{14}O$) is very close to zero, despite the 
fact that all the quantum numbers are right for an allowed Gamow-Teller 
transition. Of particular interest is the early work of Inglis \cite{inglis} 
who showed that in the simplest shell model space (2 holes in the $0p$ shell), 
it is $not$ possible to get this $GT$ matrix element to vanish if the residual 
nucleon-nucleon ($NN$) interaction consists of only a central part and a 
spin-orbit part. Inglis then commented upon the possibility that $A(GT)$ might 
vanish with only these two interactions if higher shells were included. He himself 
did not carry out such a calculation, but an attempt to do so was made a few years later 
by Baranger and Meshkov \cite{bar}. They concluded that it was possible 
that configuration mixing was the sole agent to cause $A(GT)$ to vanish; however, 
they had to speculate on the signs of certain matrix elements. 

Following Inglis' work, Jancovici and Talmi \cite{talmi} showed that if one also had a 
tensor component present in the interaction one $could$ get the $GT$ matrix 
element to vanish. Thus, the $A=14$ system affords us one of the few instances 
where one can study the elusive effects of the tensor interaction in nuclear 
structure \cite{zheng,prep}. Visscher and Ferrel \cite{viss} plotted the 
strength parameters of the spin-orbit and tensor interactions for which one 
could get the $GT$ matrix element to vanish. They noted that if the spin-orbit 
interaction is too weak then they cannot get the $GT$ matrix element to vanish 
for $any$ value of the tensor interaction strength parameter.

Zamick showed, in Ref. \cite{larry}, that the $GT$ matrix element comes out 
too large when one uses the non-relativistic $G-$matrix elements which Kuo 
\cite{kuo} obtained from the Hamada-Johnston interaction \cite{hamada}. 
However, if the spin-orbit interaction was increased by about 50\%, he could 
get this matrix element to vanish. 

Whereas most early calculations were carried out in the model space of two holes in 
the $0p$ shell, more recently, Fayache, Zamick and M\"{u}ther have reconsidered this issue 
by performing no-core shell model calculations ($NCSM$) in the larger model 
space ($[(0p)^{-2}]$ + 2$\hbar \omega$)\cite{a14}. First they used an 
interaction previously constructed by Zamick and Zheng \cite{ann}: 

\begin{equation}
V_{zz} = V_{c}+xV_{so}+yV_{t},  \label{eq1}
\end{equation}
where $c$=central, $so$=two-body spin-orbit, and $t$=tensor. For $x$=$y$=1,
the matrix elements of $V_{zz}$ are in approximate agreement with
those of the non-relativistic OBE potential Bonn A of \cite{mach2}. They then 
studied the effects of the spin-orbit and tensor components of the $NN$ 
interaction on the $GT$ matrix element by varying the strength parameters $x$ 
and $y$. They found the interesting result that in the small model space (2 holes 
in the $0p$ shell) they could (for the standard value $x=1$ of the spin-orbit 
interaction strength parameter) find a value of $y$ for which the 
$GT$ matrix element vanishes. However, in a larger model space which also 
included 2$\hbar \omega$ excitations and still using the standard value $x=1$, they 
could not get the $GT$ matrix element to vanish for $any$ value of the tensor interaction strength 
parameter $y$. Thus they reached the opposite conclusion to that of Baranger and Meshkov. 
However, if the spin-orbit interaction was enhanced by 50\% (to 
$x=1.5$), then they could find a reasonable value of $y$ for which the 
$GT$ matrix element vanishes (but not for $y=0$). Furthermore, using a relativistic Bonn A 
$G-$matrix with a Dirac effective mass $m_D=0.6m$ ($m$ being the mass of the 
free nucleon), they found that they can make the $GT$ matrix element  
vanish in both the small and large model spaces. It is known that the 
spin-orbit interaction gets enhanced by a factor $m/m_D$ in relativistic 
calculations \cite{dirac}.

In the above discussion, we have focused on the ground-state-to-ground-state 
transition $^{14}N~(J^{\pi}=1^+_1,~T=0)~\rightarrow~^{14}C~(J^{\pi}=0^+_1,
~T=1)$. But in a proposed experiment ($^{14}N(d,{^2He})^{14}C$) 
\cite{De Frenne}, one can reach excited $T=1$ states as well with spins 
$J^{\pi}=0^+,~1^+$ and $2^+$ via the allowed $GT$ mechanism. In section II, 
we shall present the results of theoretical calculations of the $GT$ reduced 
transition probability 

\begin{equation}
B(GT)=(\frac{g_A}{g_V})^2\frac{1}{2J_i+1}|A(GT)|^2,  
\end{equation}

\noindent as well as the summed strengths $\sum B(GT)$ to these states. In 
Eq. 2, $\frac{g_A}{g_V}=1.251$ is the ratio of the Gamow-Teller 
to Fermi coupling constants introduced here for convenience \cite{gagv}. 
The $GT$ matrix element itself, denoted as $A(GT)$, is given by the expression 

\begin{equation}
|A(GT)|^2=\sum_{M_i,M_f,\mu}\langle \psi_f^{J_f,M_f,T_f,T_{fz}} 
| \sum_{k=1}^A \sigma_{\mu}(k) t_+(k)~| \psi_i^{J_i,M_i,T_i,T_{iz}} \rangle^2. 
\end{equation}

We perform our calculations with a variety of realistic interactions in both 
the small and large model spaces. In light of the fact that in Ref. \cite{a14} there were 
such drastic differences between the results obtained in the small and large model spaces for 
the $GT$ transition from $J^{\pi}=1_1^+$ to $J^{\pi}=0_1^+$, we should also investigate the 
effects of going from small to large model spaces for the other allowed transitions, namely 
from $J^{\pi}=1_1^+$ to $J^{\pi}=1^+$ and to $J^{\pi}=2^+$. This is one of the main points 
of the present work. Furthermore, we will do these calculations using both a phenomenological 
approach as in \cite{a14} and a purely theoretical one in which a modern realistic $N-N$ 
effective interaction is used in larger and larger model spaces. 

The results of our calculations are presented in section II. Section III 
deals with the interpretation of the results, followed by concluding remarks.

\section{Results of the Calculations}

\indent We first show in Table I results of calculations of the $GT$ 
ground-state-to-ground-state transition strength $B(GT):~^{14}N~(J^{\pi}=
1^+_1,~T=0)~\rightarrow~^{14}C~(J^{\pi}=0^+_1,~T=1)$, followed by 
the $summed$ strengths of the $GT$ transition from the ground state of 
$^{14}N$ to the $J^{\pi}=0^+,~1^+$ and $2^+$ ($T=1$) states of $^{14}C$, 
using the interaction $V_{zz}$ (Eq. 1) of Zamick and Zheng \cite{ann}. Note 
that we never introduce a 
single-particle spin-orbit term, since in our case the average one-body 
spin-orbit interaction is implicitly generated by our two-body spin-orbit 
interaction $xV_{so}$ in our no-core shell-model ($NCSM$) calculations, which 
we performed using the nuclear shell model code $OXBASH$ \cite{oxbash}.

First, let us compare the small- and large-space results for $J^{\pi}=0^+$ 
final states obtained with the standard two-body spin-orbit strength $x=1$. 
As we vary the 
strength parameter $y$ of the tensor interaction in the small model space 
calculation, we see that for $y=0.5$ $B(GT)$ becomes vanishingly small. 
Indeed, for $y=1.0~x=1$ $A(GT)$ has an opposite sign to that for $y=0~x=1$. 
This verifies the contention of Jancovici and Talmi that one can get $A(GT)$ 
to vanish with a suitable tensor interaction. 

However, when we go to the $large$ (0+2)$\hbar \omega$ model space, we see 
that $B(GT)$ for $J^{\pi}=0^+_1$ does $not$ go to zero for any value of $y$ when 
$x=1$, and we no longer get the Jancovici-Talmi behaviour. 

The situation is restored if a combination of a weaker strength of the tensor 
interaction and an enhanced strength of the spin-orbit interaction 
($i.e.~x=1.5$ and $y=0.75$) is applied. In that case we get $B(GT)$ to vanish in 
both the $small$ and $large$ model spaces. 

We next come to one of the main points of the paper: a comparison of the $GT$ 
summed strengths to the $J^{\pi}=1^+$ and $J^{\pi}=2^+$ ($T=1$) final states 
in $^{14}C$. We see consistently that the excitation strengths to the 
$J^{\pi}=2^+$ states are much larger than to the $J^{\pi}=1^+$ states. For 
example, in the large space with $x=1.5~y=0.75$, the values of the summed 
strength to the $1^+$ states is only 0.193, but to the $2^+$ states it is 
3.113. We will discuss this further in the next section.

Table II presents results of calculations done with the relativistic Bonn A 
interaction of M\"{u}ther $et.~al.$ \cite{dirac}. In this approach, one has a 
Dirac effective mass $m_D$ such that $m_D/m$ is typically less than one  
with $m_D/m=1$ corresponding to the non-relativistic limit. In our case, the 
value of $m_D/m=0.6$ seems to work best in as far as achieving a vanishing 
$GT$ transition between the ground states of the $A=14$ system. This is true 
in both the small and the large model spaces. 

In Table III, we present the results of calculations done with the Argonne V8' 
effective interaction in four model spaces: 0$\hbar \omega$, 
(0+2)$\hbar \omega$, (0+2+4)$\hbar \omega$ and (0+2+4+6)$\hbar \omega$, all 
performed with the Many-Fermion Dynamics code of \cite{MFD}. For this set of 
calculations, we followed the 
procedure described in Refs \cite{Petr96,PRLC00} in order to 
construct the  two-body effective interaction. 
Note that in Tables I-III we give the ground-state-to-ground-state 
transition $B(GT):~^{14}N~(J^{\pi}=1^+_1,~T=0)~\rightarrow~^{14}C~(J^{\pi}=0^+_1,~T=1)$, 
as well as the summed strength $\sum B(GT)$, $i.e.$ summing the $B(GT)$ values starting from the 
$^{14}N~(J^{\pi}=1^+_1,~T=0)$ ground state to all final $0^+,~1^+$ and $2^+$ 
$T=1$ states in $^{14}C$. 
In the smallest model space, the Argonne V8' interaction gives a poor result 
for the ground-state-to-ground-state $B(GT)$, a value of 2.518 which is far 
from the desired result of $zero$. When the model space is enlarged to 
(0+2)$\hbar \omega$, the $B(GT)$ to the $0^+_1$ state obtained with the 
Argonne V8' interaction goes down to 1.403, then it goes further down to 0.430 
in the larger model space (0+2+4)$\hbar \omega$, and in the yet larger model 
space (0+2+4+6)$\hbar \omega$ it goes way down to 0.164. This shows that the 
results are quite sensitive to the model space used, but overall they are 
rather encouraging in the sense that one seems to be converging to the 
desired result that the ground-state-to-ground-state $B(GT)$ vanishes in the 
limit that the model space becomes sufficiently large. Indeed, it is clear 
that the many-body correlations in the large model spaces are causing the 
decrease in the transitions to the $0^+$ states and their increase for $2^+$ 
states. 

\section{Interpretation of the Results}

\subsection{The $L-S$ picture}

\indent We can make sense of the results obtained in the small model space 
0 $\hbar \omega$ by following the approach of Zheng and Zamick \cite{ann} 
and use an $LS$ representation ($^{2S+1}L_J$) for the 
two-hole $A=14$ system. For instance, the ground state ($J^{\pi}=1_1^+,~T=0$) 
wavefunction of $^{14}N$ ($i.e.$ the initial state) is represented as follows:

\begin{equation}
\psi_i = C_i^S\; |{}^3S_1\rangle + C_i^P\; |{}^1P_1\rangle\;
                                   + C_i^D\; |{}^3D_1\rangle \;, 
\end{equation}

\noindent whereas for final $J^{\pi}=0^+,~T=1$ states the wavefunctions are 
of the form 

\begin{equation}
\psi_f = C_f^S\; |{}^1S_0\rangle + C_f^P\; |{}^3P_0\rangle\;.
\end{equation}

\noindent The expression for the transition amplitude $A(GT)$ (see Eq. 3) is 
then 

\begin{equation}
A(GT)=\sqrt{6}[C_f^SC_i^S-C_i^PC_f^P/\sqrt{3}].
\end{equation}

It should be noted that if the $^{14}N$ ground-state 
wavefunction had a pure $^3D_1$ configuration then the Gamow-Teller transition 
amplitude $A(GT)$ to $J^{\pi}=0^+$ and $1^+$ states would vanish. The reason 
for this, of course, is that  the $GT$ operator $\sum_k \sigma_{\mu}(k) 
t_+(k)$ cannot change the orbital quantum number $L$. But from the above 
expression for $A(GT)$, it is not a necessary condition to have $C_i^D=1$ and 
$C_i^S=C_i^P=0$ in order for $A(GT)$ to vanish. Interference from the $L=1$ 
contributions can and does make $A(GT)$ vanish before $C_i^D=1$. 
Nevertheless, $C_i^D$ is very close to one at the point where $A(GT)$ 
vanishes. 

In Table IV, we present the values of the coefficients $C_i^S$, $C_i^P$ 
and $C_i^D$ related to the 0$\hbar \omega$ model space calculations done 
with various interactions considered earlier in Tables I, II and III, as 
well as the corresponding $A(GT)$. By comparing the values of the $LS$ 
coefficients shown in the upper half of this table to those in the lower one, 
it becomes clear that the argument presented above, about the 
crucial role of the $^3D_1$ component in the $J^{\pi}=1^+~T=0$ $^{14}N$ 
ground-state wavefunction in insuring the vanishing of $A(GT)$, holds for the 
other interactions as well.

We can also see why the $2^+$ final states are more strongly excited than the 
$J^{\pi}=1^+$ final states. In the two-hole model space, and by virtue of the 
generalized Pauli exclusion principle, there is only $one$ $J^{\pi}=1^+,~T=1$ 
final state, corresponding to $L=1~S=1~T=1$ and denoted by $^3P_1$. It can be 
excited by the $GT$ mechanism only via the $^1P_1$ component of the 
$^{14}N$ $J^{\pi}=1_1^+~T=0$ ground-state wavefunction. We see from Table IV 
that the $^1P_1$ component $C_i^P$ is rather small when $A(GT)$ vanishes. 
It is possible, however, to form two $J^{\pi}=2^+~T=1$ states in the 
two-hole model space (corresponding to $L=2~S=0$ and $L=1~S=1$), and the first 
one of these two configurations ($^1D_2$) will carry most of the strength of 
the $GT$ excitation emanating from the dominantly $^3D_1$ $^{14}N$ 
ground-state wavefunction ($C_i^D \ge 0.96$ when $A(GT)$ vanishes).

\subsection{Renormalization of the spin-orbit interaction}

As mentioned in the introduction, it has been noted in the past 
\cite{viss,larry,a14,ann} that the vanishing of the ground-state-to-ground 
state $GT$ matrix element in the $A=14$ system as calculated in the valence 
space ($i.e.$ 0$\hbar \omega$ model space) requires either an enhancement of the 
two-body spin-orbit interaction and/or a weakening of the tensor interaction 
-see the lower half of table IV. In a different but somewhat related context, 
Fayache $et.~al.$ had come to a similar conclusion in their study of $M1$ 
excitation rates in the $0p$ and $1d-0d$ shells \cite{npa}. It is useful to 
note here, as pointed out by Wong \cite{wong}, that the tensor interaction in 
an {\em open-shell} acts to some extent like a spin-orbit interaction of the 
{\em opposite} sign of the basic spin-orbit interaction, so that these two 
types of adjustments to the $N-N$ interaction are really equivalent for our 
purpose. 

In the present work, we have shown in table III that, using a modern 
realistic effective $N-N$ interaction and performing no-core shell-model 
calculations with it in progressively larger and larger model spaces, we were 
able to obtain the desired vanishing of the ground-state-to-ground-state $GT$ 
matrix element in a natural way, $i.e.$ without having to adjust any 
parameters. This suggests that some renormalization of the effective 
spin-orbit interaction coupling strength (in the sense of an enhancement of 
the latter relative to its strength in the 0$\hbar \omega$ model space) must 
be taking place as one works in larger and larger model spaces. 

In Table V, we present results of calculations that further corroborate the 
interpretation just given. Loosely speaking, the $J^{\pi}=2^+_1$ state is 
mainly a $(p^{-1}_{1/2})(p^{-1}_{3/2})$ two-hole state, so that its 
excitation energy scales mainly as the spin-orbit splitting 
$E({3/2}^-)-E({1/2}^-)$ in the $A=15$ system. Evidently, the latter can be 
thought of as a measure of the strength of the effective two-body spin-orbit 
interaction as calculated in a given model space. A striking systematics  
then emerges when one combines the results of tables III and V. Clearly, 
there is a one-to-one correlation between the re-distribution of the $GT$ 
strength in the $A=14$ system (table III) and the effective spin-orbit 
strength as the size of the shell-model space is varied (table V). Indeed, 
there is a clear trend taking place in the sense that, as the size of the 
shell-model space gets larger, the calculated excitation energy $E_x(2^+_1)$ 
in $^{14}C$ as well as the calculated energy splitting $E({3/2}^-)-
E({1/2}^-)$ in $^{15}N$ are increasing, and all the while a re-distribution 
of the $A=14$ $GT$ strength is taking place, with all the combined results 
becoming in better agreement with experiment.

\section{Concluding Remarks}

\indent We have performed theoretical calculations of the allowed Gamow-Teller 
transitions from the ground state of $^{14}N$ to the lowest lying states 
in $^{14}C$ in anticipation of a proposed experiment involving the reaction 
$^{14}N(d,{^2He})^{14}C$. We discussed the problem of the near vanishing of 
the $GT$ transition to the $J^{\pi}=0^+_1,~T=1$ ground state of $^{14}C$, but 
principally focused on the transitions to the final states with angular 
momenta $1^+$ and $2^+$. 

In calculations limited to a 0 $\hbar \omega$ model space, it is necessary to 
effect a phenomenological enhancement of the $N-N$ two-body spin-orbit 
interaction in order to obtain a vanishing ground-state-to-ground-state $GT$ 
matrix element in the $A=14$ system. We have found that this in turn results 
in the $GT$ strength going overwhelmingly to the lowest $2^+$ state. Such a 
result can be easily accounted for by the fact that the $^{14}N$ $J^{\pi}=
1_1^+~T=0$ ground state wavefunction is predominantly composed of an $LS$ 
component $^3D_1$. 

Using an effective interaction theoretically derived from the realistic $N-N$ 
Argonne V8' interaction, and performing shell-model calculations in 
progressively larger and larger model spaces (with up to 6 $\hbar \omega$ 
excitations), we were able to achieve a similar degree of success in 
agreement with experiment as that obtained phenomenologically earlier in the 
0$\hbar \omega$ model space calculations, but this time without any 
adjustments of parameters. We have interpreted this as an indication of a 
natural renormalization of the effective two-body spin-orbit interaction 
affected by the many-body correlations taking place in the larger 
model spaces. 

In concluding, we note that, when only the charge-symmetry-conserving strong 
interactions are taken into account (as we did in this paper), the matrix 
element for the transition from $^{14}N$ to $^{14}C$ is the same as that from 
$^{14}N$ to $^{14}O$. However, since the transitions to the $J=0^+$ states are 
strongly suppressed, large charge-symmetry effects can be induced by the 
Coulomb interaction. Indeed the $ft$ values in the decays of $^{14}C$ and 
$^{14}O$ to the ground state of $^{14}N$ are respectively $1.1~10^9$ and 
$2~10^7$, quite different values indeed. Talmi \cite{talmi2} was able to 
verify that indeed the Coulomb interaction could explain this difference to a 
large extent. It will be interesting in the near future to see the effects of 
other charge-symmetry-breaking interactions.

\section{Acknowledgements}
K. Heyde thanks D. De Frenne for
discussions on the relevance of the $(d,{^2He})$ reaction in order to
study the Gamow-Teller strength in light nuclei. L. Zamick ackowledges support 
by DOE Grant No. DE-FG02-95ER-40940. M.S. Fayache and B.R. Barrett acknowledge 
partial support from NSF Grants No. PHY0070858 and INT-0096785. B.R. Barrett 
thanks I. Talmi for useful discussions. 
This work was performed in part under the auspices of
the U. S. Department of Energy by the University of California,
Lawrence Livermore National Laboratory under contract
No. W-7405-Eng-48. P.Navratil received support from LDRD contract 00-ERD-028.
N. Smirnova and 
K. Heyde thank the DWTC for the grant IUAP \#P5/07 and the "FWO-Vlaanderen" for 
financial support. Finally, M.S. Fayache is grateful for the hospitality of the 
High-Energy Section of the Abdus Salam International Centre for Theoretical 
Physics, where part of this work was done. 

\pagebreak

\pagebreak

\begin{table}
\caption{Ground-state-to-ground-state Gamow-Teller strength $B(GT)$ (denoted by 
$0^+_1$ in columns three and seven) and summed strengths ($\sum B(GT)$) from the 
$J^{\pi}=1^+_1~T=0$ ground state of $^{14}N$ to the $J^{\pi}=0^+,~1^+$ and 
$2^+$ states in $^{14}C$, using the ($x,y$) interaction.}
\begin{tabular}{|cc|cccc|cccc|}
\hline
  & & \multicolumn{4}{c|}{ 0 $\hbar \omega$ Model Space} & 
\multicolumn{4}{c|}{ (0+2)$\hbar \omega$ Model Space} \\
\hline
 $x$    & $y$  & $0^+_1$ & $0^+$ & $1^+$ & $2^+$ & $0^+_1$ & $0^+$ & $1^+$ & $2^+$ \\
\hline
        &      &       &       &       &       &       &       &       &\\
 1      & 0    & 0.738 & 0.792 & 0.280 & 2.617 & 1.235 & 1.345 & 0.276 & 2.030\\
 1      & 0.5  & 0.002 & 0.092 & 0.148 & 3.187 & 1.682 & 2.158 & 0.009 & 0.933\\
 1      & 0.75 & 0.668 & 1.047 & 0.017 & 2.100 & 1.555 & 1.985 & 0.005 & 1.091\\
 1      & 1.0  & 0.991 & 1.427 & 0.003 & 1.707 & 1.532 & 1.929 & 0.004 & 1.132\\
        &      &       &       &       &       &       &       &       &\\
1.5     & 0    & 0.359 & 0.364 & 0.345 & 3.111 & 0.448 & 0.456 & 0.351 & 2.970\\
1.5	& 0.5  & 0.161 & 0.169 & 0.317 & 3.278 & 0.093 & 0.119 & 0.287 & 3.230\\
1.5	& 0.75 & 0.062 & 0.099 & 0.281 & 3.312 & 0.003 & 0.126 & 0.193 & 3.113\\
1.5	& 1.0  & 0.004 & 0.097 & 0.230 & 3.263 & 0.165 & 0.422 & 0.103 & 2.709\\
\hline
\hline
\end{tabular}
\end{table}

\begin{table}
\caption{Same as Table I, but using M\"{u}ther's relativistic Bonn A
interaction, characterized by the ratio of the nucleon's Dirac mass $m_D$ 
to its free mass $m$.}

\begin{tabular}{|c|cccc|cccc|}
\hline
          & \multicolumn{4}{c|}{ 0 $\hbar \omega$ Model Space} & 
\multicolumn{4}{c|}{ (0+2)$\hbar \omega$ Model Space} \\
\hline
 $m_D/m$  & $0^+_1$ & $0^+$ & $1^+$ & $2^+$ & $0^+_1$ & $0^+$ & $1^+$ & $2^+$ \\
\hline
       &        &       &       &       &       &       &       &\\
  1.0  &  1.765 & 2.144 & 0.004 & 0.990 & 0.779 & 1.064 & 0.013 & 2.001\\
       &        &       &       &       &       &       &       &\\
  0.75 &  0.051 & 0.194 & 0.124 & 3.061 & 0.043 & 0.168 & 0.119 & 2.986\\
       &        &       &       &       &       &       &       &\\
  0.60 &  0.033 & 0.085 & 0.255 & 3.300 & 0.001 & 0.078 & 0.180 & 3.115\\
\hline
\hline
\end{tabular}
\end{table}

\begin{table}
\caption{Same as Table I, but using 
two-body effective interactions derived from the Argonne V8'
NN potential without Coulomb. The HO frequency of $\hbar\omega=14$ MeV 
was employed. The calculated $6\hbar\omega$ binding energies of $^{14}$N is
110.52 MeV. The binding energy is expected to decrease with a further
enlargement of the basis size.}
\begin{tabular}{|c|cccc|}
\hline
  Model Space & $0^+_1$ & $0^+$ & $1^+$ & $2^+$ \\
\hline
       &        &        &       &       \\
   0 $\hbar \omega$      & 2.518 & 2.905 & 0.0262 & 0.2511 \\
       &        &        &       & \\
 (0+2)$\hbar \omega$     & 1.403 & 1.839 & 0.0004 & 1.230 \\
       &        &        &       & \\
 (0+2+4)$\hbar \omega$   & 0.430 & 0.799 & 0.0291 & 2.264 \\
       &        &        &       & \\
 (0+2+4+6)$\hbar \omega$ & 0.164 & 0.480 & 0.318 & 3.081  \\
\hline
\hline
\end{tabular}
\end{table}

\begin{table}
\caption{The $LS$-representation coefficients of the 
$J^{\pi}=1^+_1~T=0$ ground state wavefunction of $^{14}N$ (Eq. 4), and the 
Gamow-Teller amplitude $A(GT)$ to the $J^{\pi}=0_1^+~T=1$ ground state of 
$^{14}C$ (Eq. 6) for various interactions considered in the previous tables.
For the AV8$^\prime$ interaction, the $0\hbar\omega$ basis results are shown.}
\begin{tabular}{|c|ccc|c|}
\hline
 Interaction            & $C_i^S$ & $C_i^P$ & $C_i^D$ & $A(GT)$\\
\hline
 $x$=1.0, $y$=1.0       &  0.675  &  0.032  &  0.737  & 1.378 \\
                        &         &         &         &       \\
 Bonn A ($m_D/m=1$)     &  0.827  & -0.037  &  0.561  & 1.839 \\
                        &         &         &         &       \\
 Argonne $V8'$          &  0.963  & -0.093  &  0.255  & 2.197 \\
                        &         &         &         &       \\
\hline
 $x$=1.0, $y$=0.49      &  0.086  &  0.224  &  0.971  & 0.000 \\
                        &         &         &         &       \\
 $x$=1.44, $y$=1.0      &  0.116  &  0.256  &  0.960  & 0.000 \\
                        &         &         &         &       \\ 
 Bonn A ($m_D/m=0.6$)   &  0.057  &  0.364  &  0.930  & 0.250 \\
\hline
\hline
\end{tabular}
\end{table}

\begin{table}
\caption{Calculated excitation energy of the $2^+_1$ state in the $A=14$ 
system and calculated spin-orbit splitting in the $A=15$ system (in MeV) in various 
model spaces, using two-body effective interactions derived from the Argonne V8'
NN potential without Coulomb. The HO frequency of $\hbar\omega=14$ MeV was used.
The calculated $6\hbar\omega$ binding energies are 108.65 MeV and 126.73 MeV 
for $^{14}$C and $^{15}$N, respectively. The binding energy is expected 
to decrease with a further enlargement of the basis size. The experimental values of 
$E_x(2^+_1)$ in $^{14}O$ and of $E({3/2}^-)-E({1/2}^-)$ in $^{15}O$ given under Expt 
are from Nuclear Data Retrieval ($http://www.nndc.bnl.gov$).}

\begin{tabular}{|c|cc||cc|}
\hline
  Model Space & $A=14$ $E_x(2^+_1)$ & Expt & $A=15$ $E({3/2}^-)-E({1/2}^-)$ & Expt \\
\hline
       &        &       &       &        \\
   0 $\hbar \omega$      & 3.152 & 6.59  & 3.314 & 6.176 \\
       &        &       &       &          \\
 (0+2)$\hbar \omega$     & 4.854 &  ''     & 5.366 & '' \\
       &        &       &       &          \\
 (0+2+4)$\hbar \omega$   & 5.564 &  ''   & 6.326 & '' \\
       &        &       &       &          \\
 (0+2+4+6)$\hbar \omega$ & 5.874 &  ''   & 6.731 & '' \\
\hline
\hline
\end{tabular}
\end{table}

\end{document}